\documentclass[useAMS,usenatbib,usegraphicx]{mn2e}
\usepackage{bm}
\usepackage{times} 
\def\sdss{SDSS J1048+4637}

\begin{document}
\title[Flattened extinction curves]{Extinction curves
flattened by reverse shocks in supernovae}
\author[H. Hirashita et al.]{Hiroyuki Hirashita$^{1}$\thanks{E-mail:
     hirasita@ccs.tsukuba.ac.jp}, %%\thanks{Postdoctoral
%%     Fellow of the Japan Society for the Promotion of Science (JSPS).}, 
Takaya Nozawa$^{2}$, Tsutomu T. Takeuchi$^{3}$, and
Takashi Kozasa$^{2}$
%%\newauthor
%%Takako T. Ishii$^{4}$\dag,
\newauthor
\\
$^1$ Centre for Computational Sciences, University of Tsukuba,
     Tsukuba 305-8577, Japan \\
$^2$ Department of Cosmosciences, Graduate
     School of Science, Hokkaido University, Sapporo
     060-0810, Japan \\
$^3$ Institute for Advanced Research, Nagoya University, Nagoya
     464-8601, Japan
}
\date{2007 December 7}
\pubyear{2007} \volume{000} \pagerange{1}
\twocolumn
%%\onecolumn

\maketitle \label{firstpage}
\begin{abstract}
We investigate the extinction curves of young galaxies in which
dust is supplied from Type II supernovae (SNe II) and/or pair
instability supernovae (PISNe). Since at high redshift ($z>5$),
low-mass stars cannot be dominant sources for dust grains,
SNe II and PISNe, whose progenitors are massive stars with
short lifetimes, should govern the dust production. Here, we
theoretically investigate the extinction curves of dust
produced by SNe II and PISNe, taking into account reverse
shock destruction induced by collision with ambient
interstellar medium. We find that the extinction curve is
sensitive to the ambient gas density around a SN, since the
efficiency of reverse shock destruction strongly depends on
it. The destruction is particularly efficient for
small-sized grains, leading to a flat extinction curve in
the optical and ultraviolet wavelengths. Such a large ambient
density as $n_{\rm H}\ga 1$ cm$^{-3}$ produces too flat an
extinction curve to be consistent with the observed extinction
curve for \sdss\ at $z=6.2$. Although the extinction curve is
highly sensitive to the ambient density, the hypothesis
that the  dust is predominantly formed by SNe at
$z\sim 6$ is still allowed by the current observational
constraints. For further quantification, the ambient density
should be obtained by some other methods. Finally we also
discuss the importance of our results for observations of
high-$z$ galaxies, stressing a possibility of flat
extinction curves.
\end{abstract}

\section{Introduction}

Dust grains play an important role in the formation and
evolution of galaxies. Dust grains control the energy
balance in the interstellar medium (ISM) by absorbing
stellar light and reemitting it in far infrared (FIR).
Also, the surface of dust grains is a site for an
efficient formation of H$_2$ molecules
\citep[e.g.][]{cazaux04}, which act as an effective
coolant in metal-poor ISM. Those effects of dust turn on
even at $\sim 1$\% of the solar metallicity according to
the calculation by \citet{hirashita02}, who argue that the
star formation rate is enhanced because of the first dust
enrichment in the history of galaxy
evolution. The first sources of dust in the Universe are
Type II (core-collapse) supernovae (SNe II) or pair
instability supernovae
(PISNe), since the lifetimes of their progenitors are short
($\sim 10^6$ yr). In the local Universe, dust grains are
also produced by evolved low mass stars \citep{gehrz89},
but this production mechanism requires much longer
($\ga 1$ Gyr) timescales. The first dust supplied by
SNe II or PISNe may trigger the
formation of low-mass stars via dust cooling
\citep{schneider03,omukai05}.

To quantify the above effects of dust in the early stages
of galaxy evolution, it is crucial to know how
much dust and what species of grains form in supernovae (SNe).
It has been suggested by some observations of nearby
SNe that dust is indeed produced in SNe, although the quantity
of formed dust is still debated
\citep[e.g.][]{moseley89,dunne03,morgan03,hines04,sugerman06,meikle07,rho07}.
By treating the nucleation and accretion in SNe,
the dust composition and size distribution are
theoretically calculated \citep*{kozasa89,kozasa91}.
Recently, in order to
examine the effects of dust in Population III
(Pop III) objects, the formation of dust in SNe II and
PISNe is extensively examined
(\citealt{todini01}; \citealt[hereafter N03]{nozawa03},
\citealt{schneider04}).
The motivation for considering PISNe comes from some
evidence indicating that the stars formed from
metal-free gas, Population III (PopIII) stars, are very
massive with a characteristic mass of a few hundred
solar masses \citep[e.g.][]{nakamura01,bromm04}.
Such massive stars are
considered to begin pair creation of electron and
positron after the helium burning phase, and finally
an explosive nuclear reaction
disrupts the whole stars \citep{fryer01,heger02}.
This explosion is called PISN.

Recently, the hypothesis that the dust formation is
dominated by SNe II and/or PISNe at high redshift ($z$) is
observationally examined for some objects.
We expect that at $z>5$, when the cosmic age is less than
1 Gyr, the main sources of dust grains are SNe II and
PISNe \citep[e.g.][]{dwek07}. Extinction
curves can be used to investigate
the dust properties \citep[e.g.][]{mathis90}. By using a
sample of broad absorption line (BAL) quasars,
\citet{maiolino04a} show that the extinction properties of
the low-$z$ ($z<4$)
sample is different from those of the high-$z$ ($z>4.9$)
sample. This result is suggestive of a
change in the dust production mechanism in the
course of galaxy evolution.
The highest-$z$ BAL quasar
in their sample, SDSS J104845.05+463718.3
(hereafter \sdss) at $z=6.2$ shows an extinction curve
flat at wavelength
$\lambda \ga 1700$ \AA\ and rising at
$\lambda\la 1700$ \AA.
\citet{maiolino04b} show
that the extinction curve
of \sdss\  is
in excellent agreement with the SN II dust models by
\citet{todini01}. \citet{bianchi07} consider dust
destruction by reverse shock in SNe,
suggesting that 2--20\% of the initial dust mass
survives, and that the extinction curve after the
destruction is still
consistent with that of \sdss.
More recently, \citet{stratta07} show that the dust
extinction in the host galaxy of GRB 050904 at
$z=6.3$ can be explained by the extinction
curve of \sdss, further supporting that the SNe II are the
main sources of dust at $z>6$.
Also, \citet{willott07}
find a similar extinction property for CFHQS J1509-1749
($z=6.12$) to that of \sdss.

There are other series of theoretical papers on
the extinction curves of high-$z$ objects.
\citet[][hereafter H05]{hirashita05} calculate the
extinction curve based on the dust production
calculation by N03. They also reproduce the extinction
curve of \sdss\ by using the
the dust production in SNe II, although
the dust composition and size distribution are
different from those of \citet{todini01}.
Recently, \citet[][hereafter N07]{nozawa07} have
treated the dust destruction by the reverse shock
as done by \citet{bianchi07}, but considering the
motion of dust relative to gas caused by the
drag force and the destruction of dust in the
radiative phase as well as in the non-radiative
phase of supernova remnants. Then, they show
that the size distribution of grains
supplied in the ISM is strongly modified by
the reverse shock.
Grains smaller than $\sim 0.02~\mu$m are
efficiently destroyed if the ambient hydrogen
number density is larger than 0.1 cm$^{-3}$.
Thus, it is important to reexamine the consistency
between the observed extinction curve and the
reverse shock destruction.

In this paper, we calculate the extinction curves
based on the dust properties calculated by N07,
who have focused on the effect of reverse shock
destruction in SNe. Then, we compare the results with
observed extinction curves at high $z$. This paper is
organized as follows. First, we
describe our theoretical treatment to calculate the
extinction curves of SN II and PISN dust in
Section \ref{sec:model}. We show and examine our
results in Section \ref{sec:results}. We discuss our
results from the
observational viewpoint in Section \ref{sec:obs}, and
finally give the conclusion of this paper in
Section \ref{sec:sum}.

\section{Model}\label{sec:model}

We derive the theoretical extinction curves of dust
grains produced in SNe II and PISNe and subsequently
destroyed by the reverse shock. Those grains are
considered to be supplied in the interstellar spaces.
The grain composition and size distribution in SNe
before the destruction is calculated by N03, whose
results are adopted as the initial conditions for the
calculations of
reverse shock destruction by N07. By using the results by
N07, the extinction curves are calculated by the same
method as in H05. The outline of our
calculation is reviewed as follows.

\subsection{Dust production and destruction in SNe II
and PISNe}\label{subsec:nozawa}

N03 calculate the dust composition and size distribution
in the ejecta of PopIII SNe II and PISNe based on the
supernova model of \citet{umeda02}, carefully
treating the radial density profile and the temperature
evolution. As
mentioned in H05, the resulting grain composition and
size distribution are
not sensitive to the metallicity of progenitor (N03).
Thus, the assumption of zero-metallicity is not
essential in this paper, and our results can be
applicable to metal-enriched systems.
Since it is still uncertain how efficiently the
mixing of atoms within SNe occurs, N03
treat two extreme cases for the mixing of elements: one
is the {\it unmixed} case in which the original onion-like
structure of elements is preserved, and the other is the
{\it mixed} case in which the elements are uniformly
mixed within the helium core. They show that the
formed dust species
depend largely on the mixing of seed elements within
SNe, because the dominant reactions change depending on
the ratio of available elements. The formed grain species
in the calculation of
N03 are listed in
Table \ref{tab:species}.

\begin{table}
\centering
\begin{minipage}{80mm}
\caption{Summary of grain species.}
\begin{tabular}{@{}lccc@{}}\hline
Species & condition\footnote{The classifications ``m'', ``u'',
and ``m/u'' mean that the
species is formed in mixed, unmixed, and both supernovae,
respectively.} & Ref\footnote{References for optical
constants: (1) \citet{edo83}; (2) \citet{piller85};
(3) \citet{philipp85}; (4) \citet{semenov03};
(5) A. Triaud and H. Mutschke (2006 private communication and see 
http://www.astro.uni-jena.de/Laboratory/OCDB/oxsul.html);
(6) \citet{toon76}; (7) \citet{roessler91};
(8) \citet{semenov03} (The optical constants of Mg$_2$SiO$_4$
are used for $\lambda\leq 0.3~\mu$m).}
&  density ($\delta_j$) \\
& & & (g cm$^{-3}$)
%%\multicolumn{2}{c}{[$M_\odot~{\rm yr}^{-1}~{\rm kpc}^{-2}$]}
 \\ \hline
C  & u & 1 & 2.28 \\
Si & u & 2 & 2.34 \\
SiO$_2$& m/u & 3 & 2.66 \\
Fe & u & 4 & 7.95 \\
FeS & u & 4 & 4.87 \\
Fe$_3$O$_4$ & m & 5 & 5.25 \\
Al$_2$O$_3$ & m/u & 6 & 4.01 \\
MgO     & u & 7 & 3.59 \\
MgSiO$_3$ & m/u & 8 & 3.20 \\
Mg$_2$SiO$_4$ & m/u & 4 & 3.23 \\
\hline
\end{tabular}
\label{tab:species}
\end{minipage}
\end{table}

In the unmixed ejecta, a variety of grain species
(Si, Fe, Mg$_2$SiO$_4$, MgSiO$_3$, MgO, Al$_2$O$_3$,
SiO$_2$, FeS, and C) condense, while in the mixed ejecta,
only oxide grains (SiO$_2$, MgSiO$_3$, Mg$_2$SiO$_4$,
Al$_2$O$_3$, and Fe$_3$O$_4$) form. The species are
summarized in Table \ref{tab:species}. Based on the
results in N03, N07 treat the
dust destruction by the reverse shock in the supernova
remnant. They find that small-sized grains suffer dust
destruction by the reverse shock and that the final
grain size distribution
is biased to larger grains than the original distribution
calculated by N03. We adopt their results as the properties
of grains supplied to the interstellar space. Following
H05, we adopt the representative progenitor mass of SNe II
as 20 $M_\odot$ and that of PISNe as 170 $M_\odot$. We
also investigate the mixed and unmixed cases.
Therefore, we treat four cases:

\noindent
(a) mixed SNe II;

\noindent
(b) unmixed SNe II;

\noindent
(c) mixed PISNe; and

\noindent
(d) unmixed PISNe.

\noindent
All the formulation and the results can be seen in
N03 and N07. The grains are assumed to be homogeneous
and spherical.

\subsection{Optical constants and extinction curves}
\label{subsec:method}

In order to calculate extinction curves, the optical
constants of grains are necessary. We adopt the
references listed in Table \ref{tab:species} for the
optical constants. Here, we only explain the difference
from H05. The optical constants are updated for
Fe$_3$O$_4$. For MgSiO$_3$, we adopt the optical
constants of Mg$_2$SiO$_4$ at $\lambda\leq 0.3~\mu$m,
where the currently available experimental data are
relatively poor \citep{dorschner95}.
For Si and SiO$_2$, we adopt the optical constants
for amorphous instead of crystal since the grains
rapidly grow in SNe. Subsequent sputtering due to
the reverse shock may also tend to destroy crystal
structures and to form more complicated amorphous
solids. The
infrared broad feature at $\lambda\simeq 21~\mu$m
is also better explained by amorphous SiO$_2$ than by
crystal SiO$_2$ \citep{rho07}. We discuss the
uncertainty in the assumed optical constants
in Appendix \ref{subsec:uncertainty}.

By using those optical constants, we calculate the
absorption and scattering cross sections of
homogeneous spherical grains with various sizes
based on the Mie theory \citep{bohren83}. Then, the
opacity of grain $j$ ($j$ denotes a grain
species) as a function of
wavelength, $\tau_{\lambda ,\, j}$, is calculated by
weighting the cross sections according to the size
distribution.
The extinction curve is presented in the form of
$A_\lambda /A_{\lambda_0}$ ($A_\lambda$ is the
extinction in
units of magnitude at wavelength $\lambda$, and
$\lambda_0$ is a reference wavelength).
The extinction in units
of magnitude is proportional to the optical depth as
$A_{\lambda,\, j}=1.086\tau_{\lambda ,\, j}$,
where $A_{\lambda,\,j}$ is the extinction of species
$j$ in units of magnitude as a function of $\lambda$.
The total extinction $A_\lambda$ is
calculated by summing $A_{\lambda,\,j}$ for all the
concerning species:
\begin{eqnarray}
A_\lambda =\sum_j A_{\lambda ,\, j}\, .
\label{eq:extall}
\end{eqnarray}
For more details, see H05. Since we normalize the
extinction at a certain wavelength $\lambda_0$,
only relative values of
$\tau_{\lambda ,\, j}$ are important.

\section{RESULTS}\label{sec:results}

\subsection{Extinction curves after the reverse
shock destruction}\label{subsec:theor}

\begin{figure*}
\includegraphics[width=8cm]{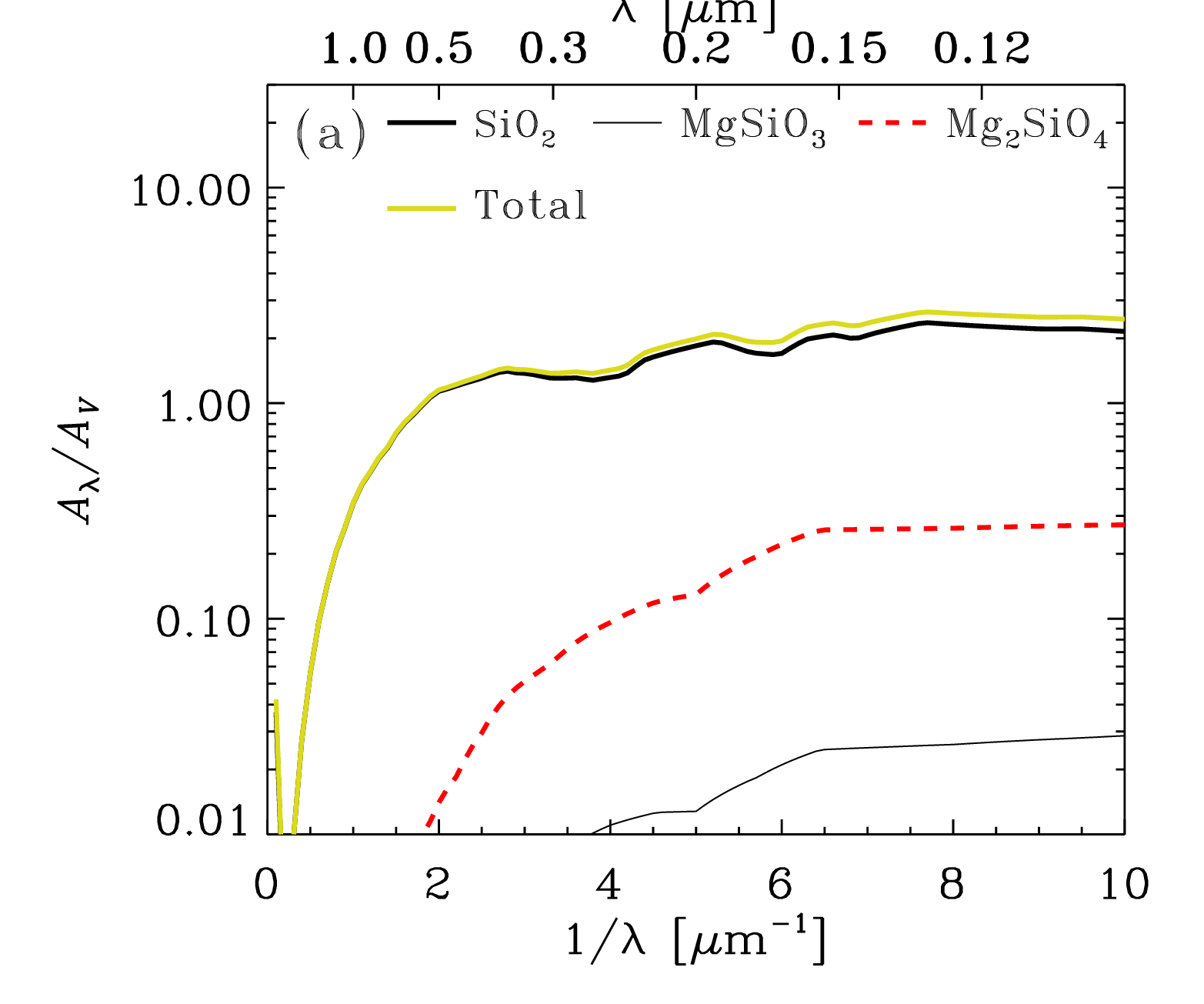}
\includegraphics[width=8cm]{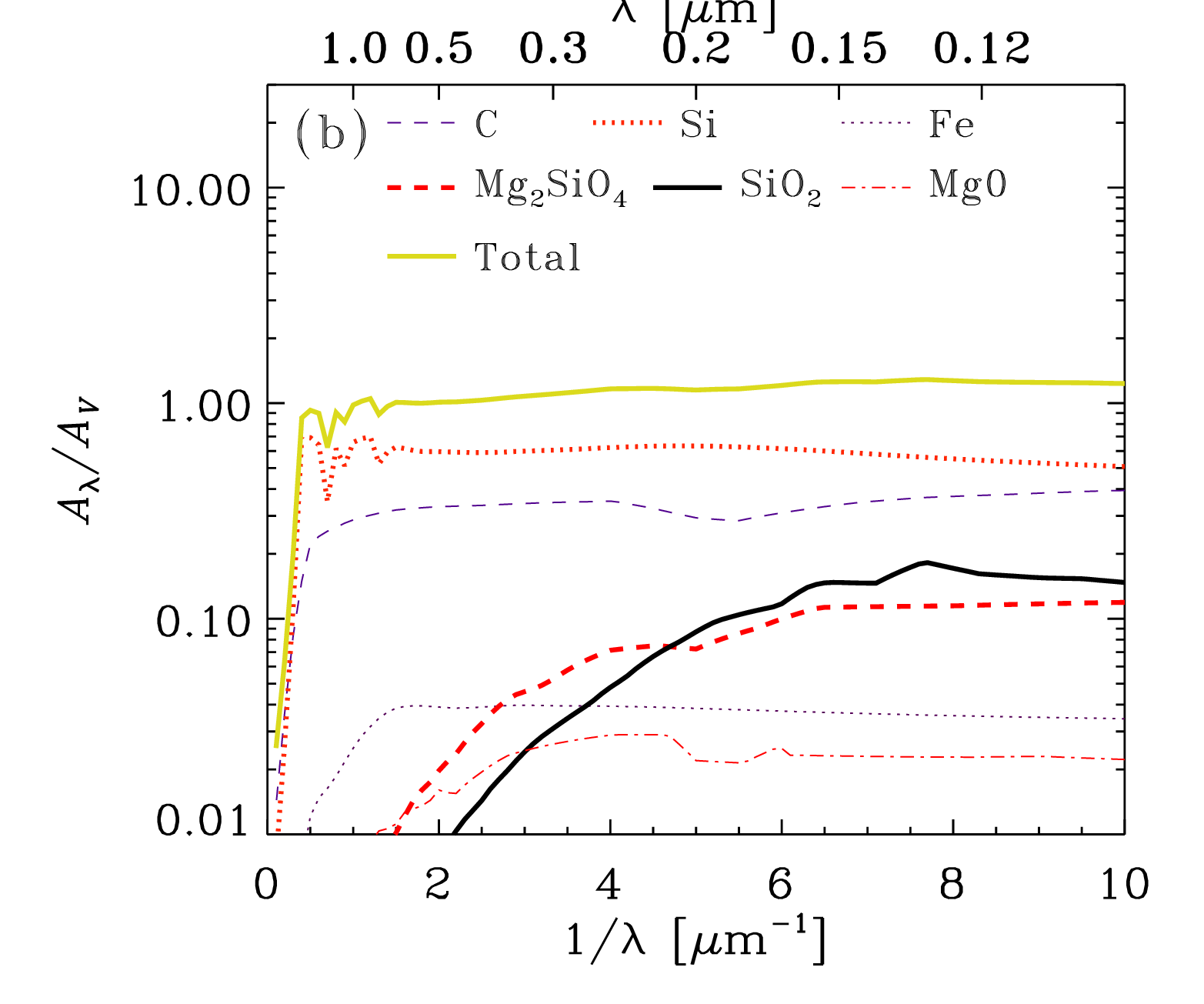}
\includegraphics[width=8cm]{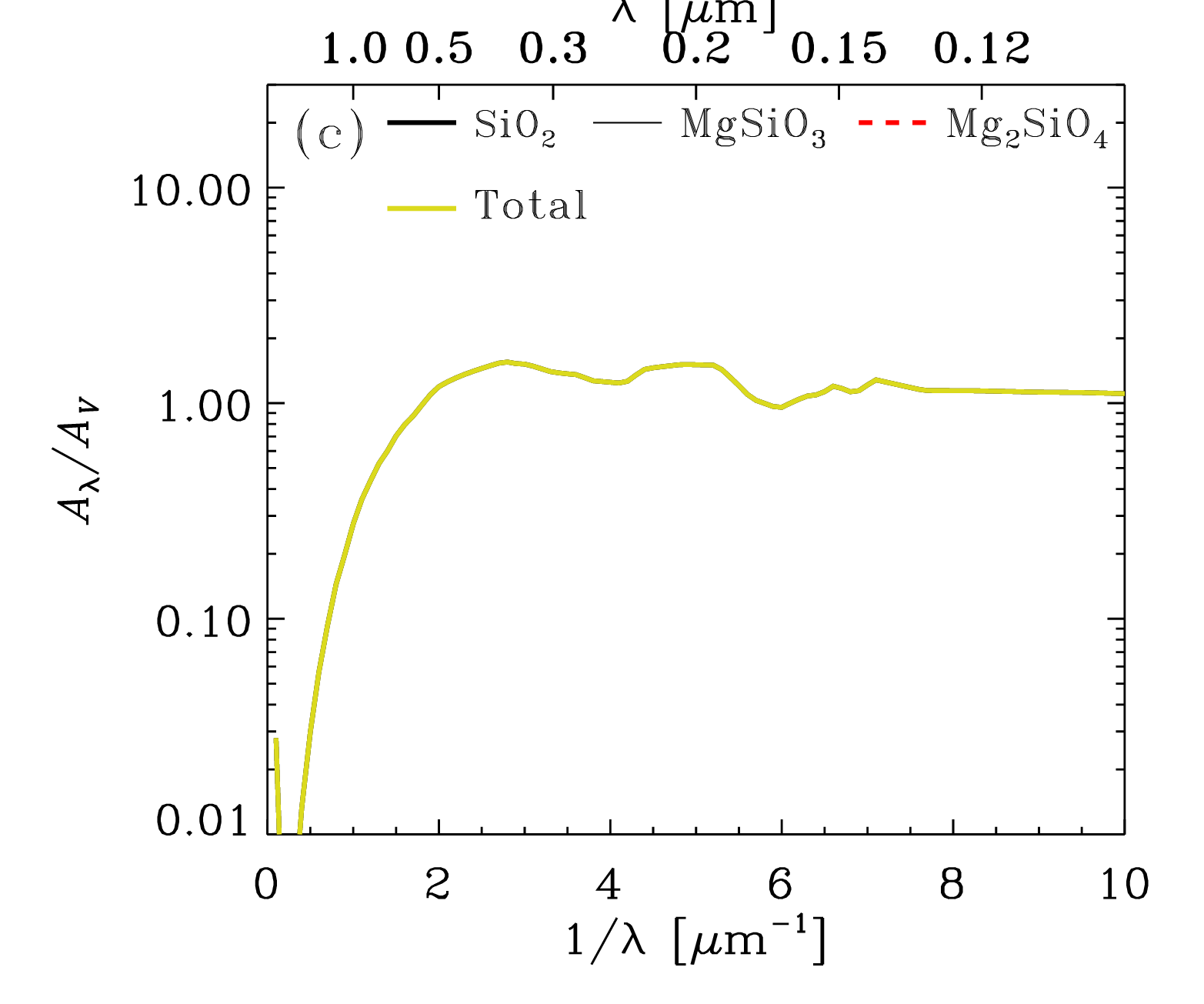}
\includegraphics[width=8cm]{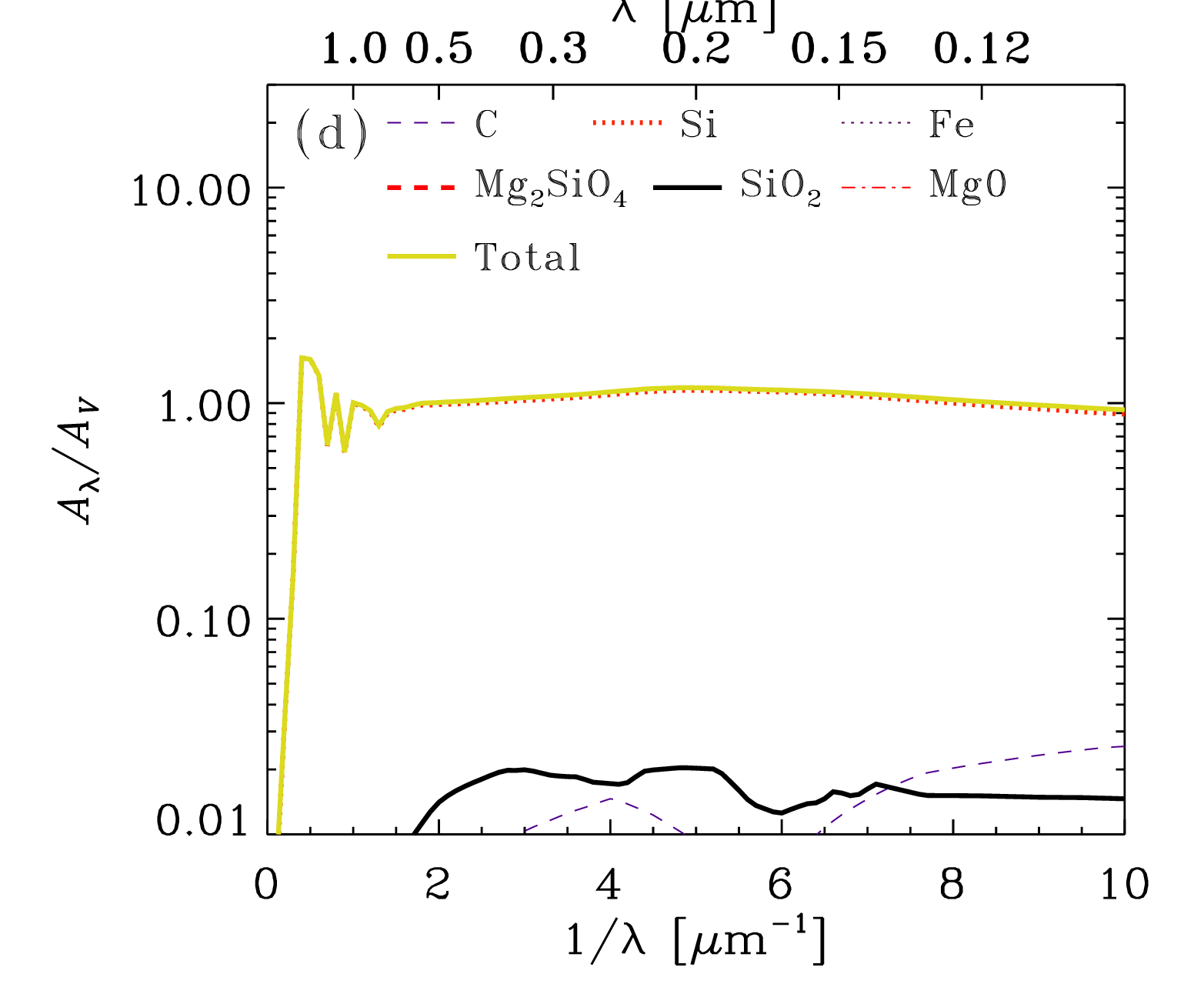}
\caption{Extinction curves of grains produced
in (a) the mixed ejecta with the progenitor of 20 $M_\odot$,
(b) the unmixed ejecta with the progenitor of 20 $M_\odot$,
(c) the mixed ejecta with the progenitor of 170 $M_\odot$,
and (d) the unmixed ejecta with the progenitor of
170 $M_\odot$. The ambient hydrogen number density
is assumed to be $n_{\rm H}=1$ cm$^{-3}$. The extinction
is normalized to the value at $\lambda =0.55~\mu$m.
The correspondence between the species and the
lines is shown in each panel. Note that some of the
species are almost completely destroyed and do not
appear on the figures and that in Panel (c), only
SiO$_2$ contributes to the total extinction curve.
\label{fig:n1}}
\end{figure*}

As shown in N07, the final size distribution of grains
after the reverse shock destruction is sensitive to the
density of the ambient ISM. Here we start from the
results for $n_{\rm H}=1$ cm$^{-3}$, where $n_{\rm H}$
is the hydrogen number density in the ambient medium,
to examine the effect of grain destruction on the
extinction curve. In Figure \ref{fig:n1}, we show the
resulting extinction
curves of the four cases in Section \ref{subsec:nozawa}.
The contribution of each species is
also shown. We normalize
the extinction to $A_V$ (i.e., $\lambda_0=0.55~\mu$m).

The extinction curves of dust produced by the mixed
SNe II and PISNe are dominated by SiO$_2$
(Figures \ref{fig:n1}a and c; in the PISN case,
only the contribution from SiO$_2$ appears and the
total extinction curve is identical to the extinction
curve of SiO$_2$). The extinction curve is flatter
than that without destruction because the reverse shock
efficiently destroys small grains and the mean grain
size becomes larger.

The extinction curve for the unmixed SNe II
(Figure \ref{fig:n1}b) is dominated by Si and C.
However, as shown in H05, the extinction curve without
reverse shock destruction is dominated by Mg$_2$SiO$_4$
and FeS for $\lambda\la 0.5~\mu$m and
by Si for $\lambda\ga 0.5~\mu$m.
Since Si and C grains are larger than the other
species, they survive the reverse shock more
than the others. For the same reason,
the extinction curve of unmixed PISNe is dominated by
Si. Also for the unmixed SNe II and PISNe,
the extinction curves of grains after the
reverse shock destruction are flatter, since
small-sized grains are selectively destroyed and the
mean size of the grains becomes large.

\subsection{Dependence on the ambient medium density}
\label{subsec:ambient}

\begin{figure*}
\includegraphics[width=8cm]{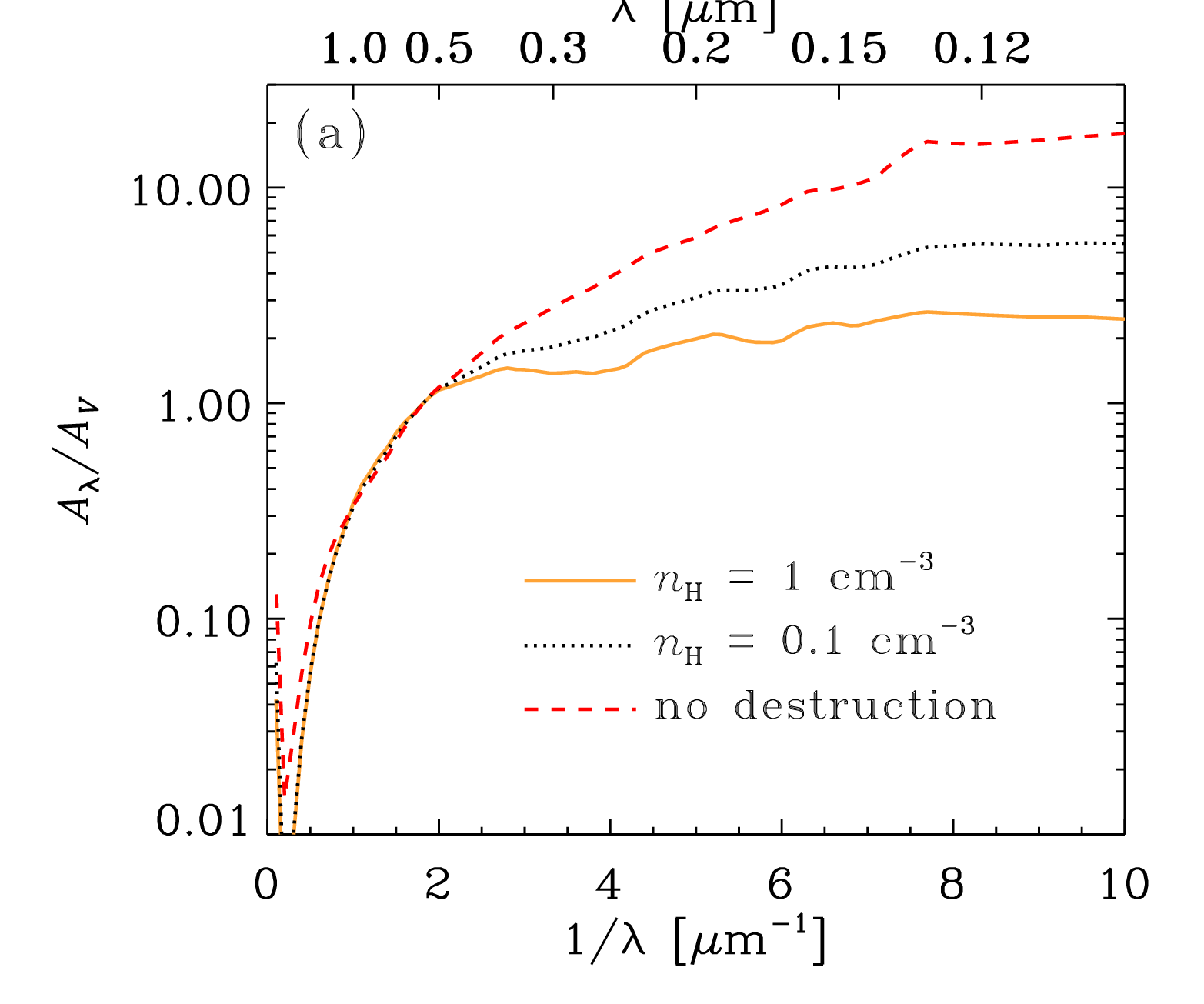}
\includegraphics[width=8cm]{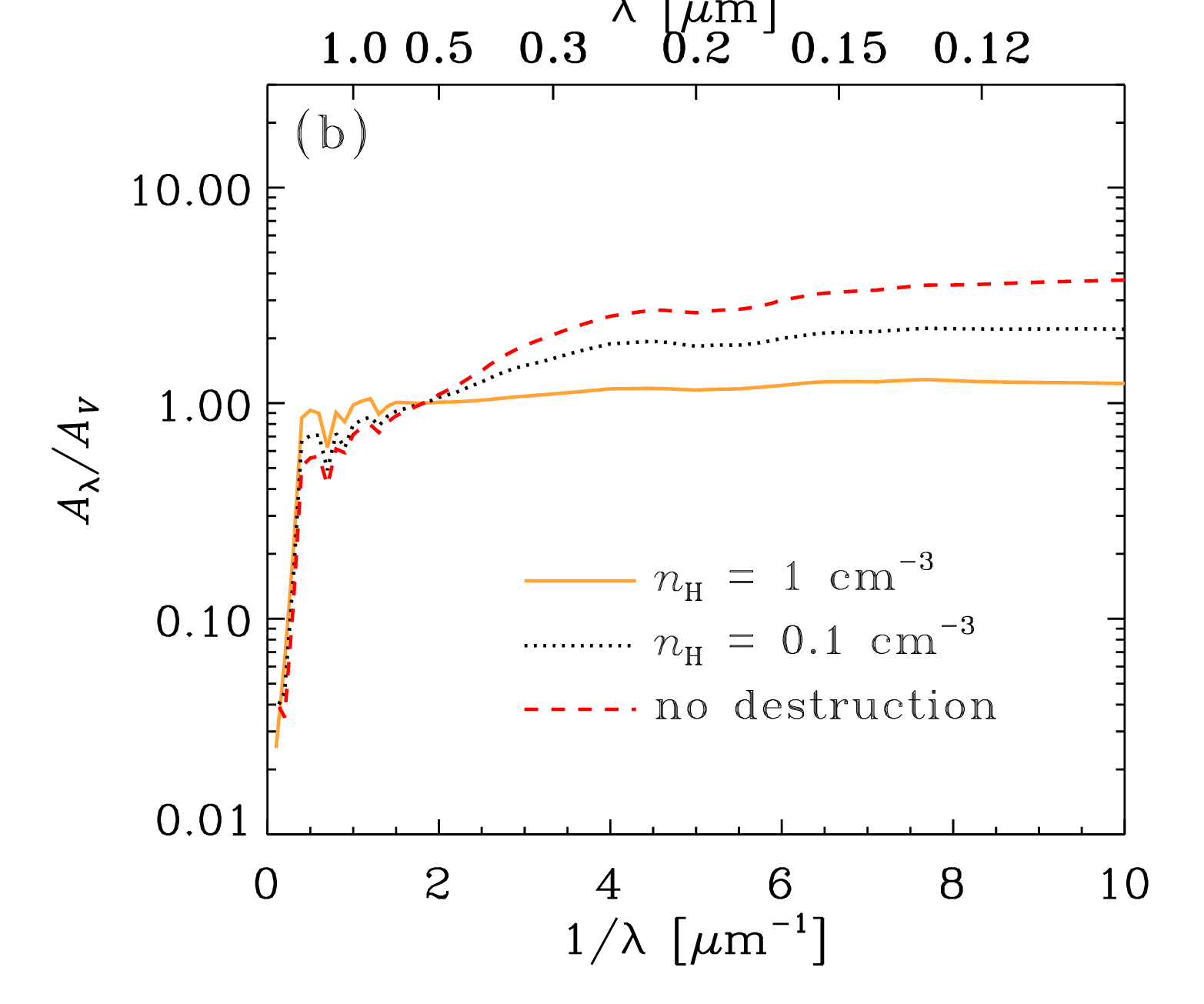}
\includegraphics[width=8cm]{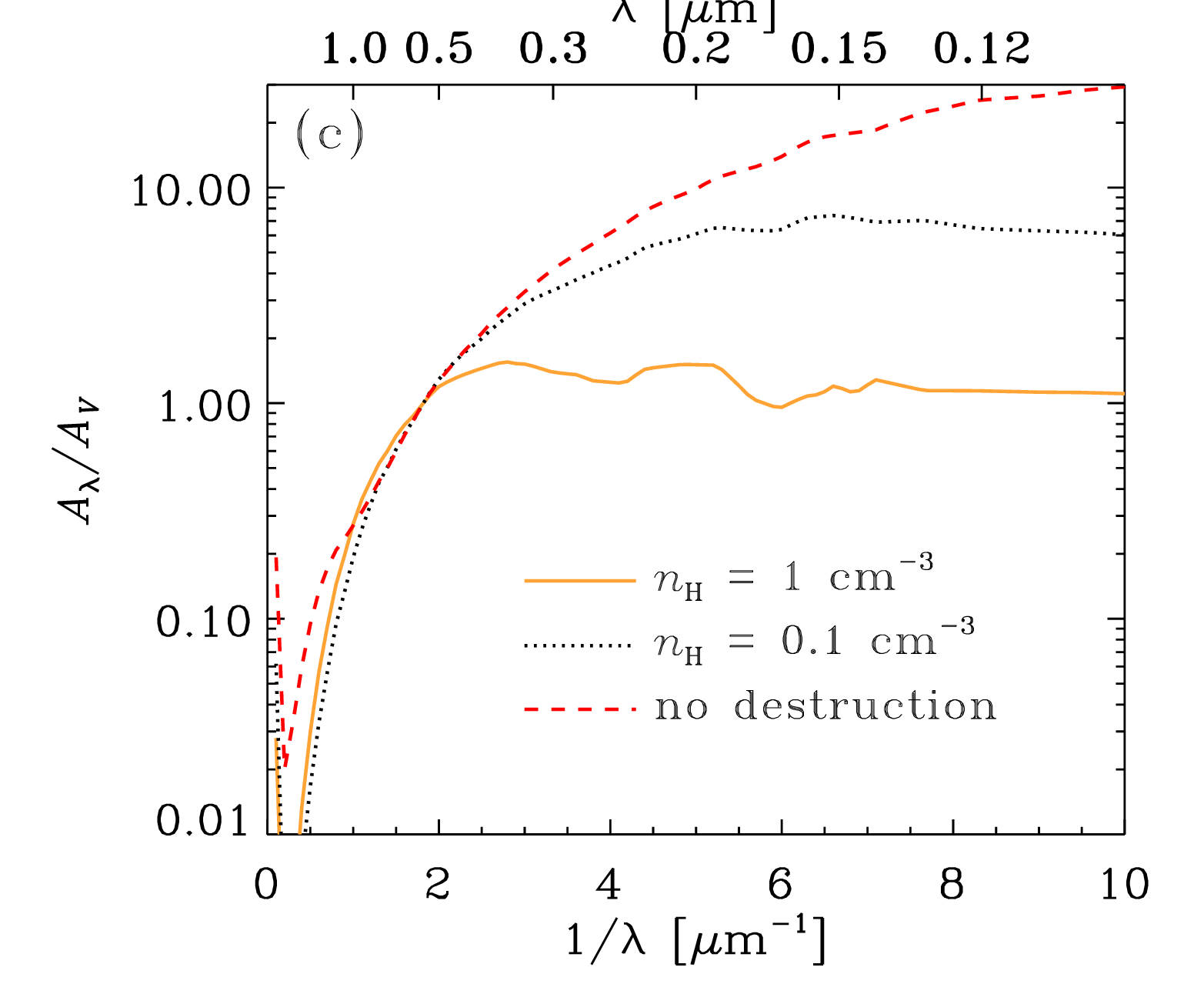}
\includegraphics[width=8cm]{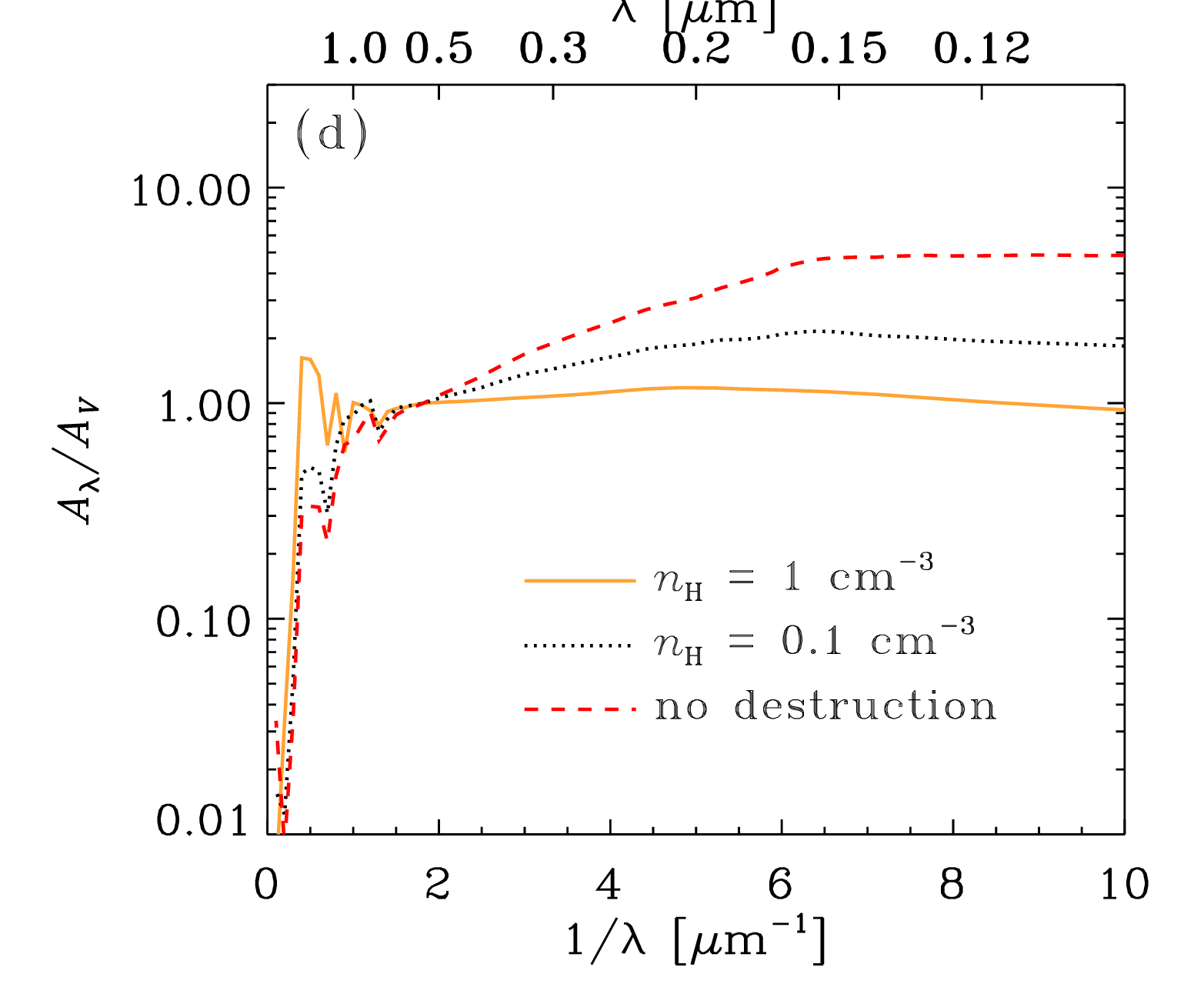}
\caption{Extinction curves of grains for various
ambient hydrogen number densities ($n_{\rm H}=1$ and
0.1 cm$^{-3}$ for the solid and dashed lines,
respectively). The case without the reverse shock
destruction (non-destruction case) is
also shown in each
panel (the dashed line). The progenitor model in
each panel is
the same as that in Figure \ref{fig:n1}.
\label{fig:ncomp}}
\end{figure*}

As shown by N07, the efficiency of the reverse shock
destruction is sensitive to the density of the ambient
medium. Here we investigate the variation of the
extinction curve for various ambient
densities. N07 have
examined three cases for the ambient hydrogen number
density: $n_{\rm H}=0.1$, 1, and 10 cm$^{-3}$.
N07 show that the grains are almost completely destroyed
for $n_{\rm H}=10$ cm$^{-3}$. Thus, we calculate
the cases for $n_{\rm H}=0.1$ and 1 cm$^{-3}$.
For comparison, we also show the results without the
reverse shock destruction (this case is called
non-destruction case). For the details for the
non-destruction case, see N03 and H05.
The results are shown in
Figure \ref{fig:ncomp}.
As expected, the extinction curve becomes flatter
for a larger ambient density because the destruction
efficiency of small-sized grains is larger.
In Section \ref{sec:obs}, our results are compared with
high-$z$ data of extinction curve.

\citet{bianchi07} also state that the extinction curve
becomes flatter, although they start from a different
dust formation model based on \citet{todini01}.
The differences between \citet{bianchi07} and the
present work are partly due to the differences in
optical constants: amorphous carbon is responsible for
most of the extinction in their work. Furthermore,
their results show more survival of small grains
than those of N07. As mentioned in N07, the difference
comes from the treatment of grain motion: In N07,
the trap of grains in a hot and dense zone between
forward and reverse shocks is properly treated, and
this effect enhances the destruction of small-sized
grains. Thus, our results show flatter extinction 
curves than those of \citet{bianchi07} especially
when the ambient gas density is large.

\section{OBSERVATIONAL DISCUSSION}\label{sec:obs}

\subsection{Comparison with high-$z$ data}
\label{subsec:comp}

\begin{figure*}
\includegraphics[width=8cm]{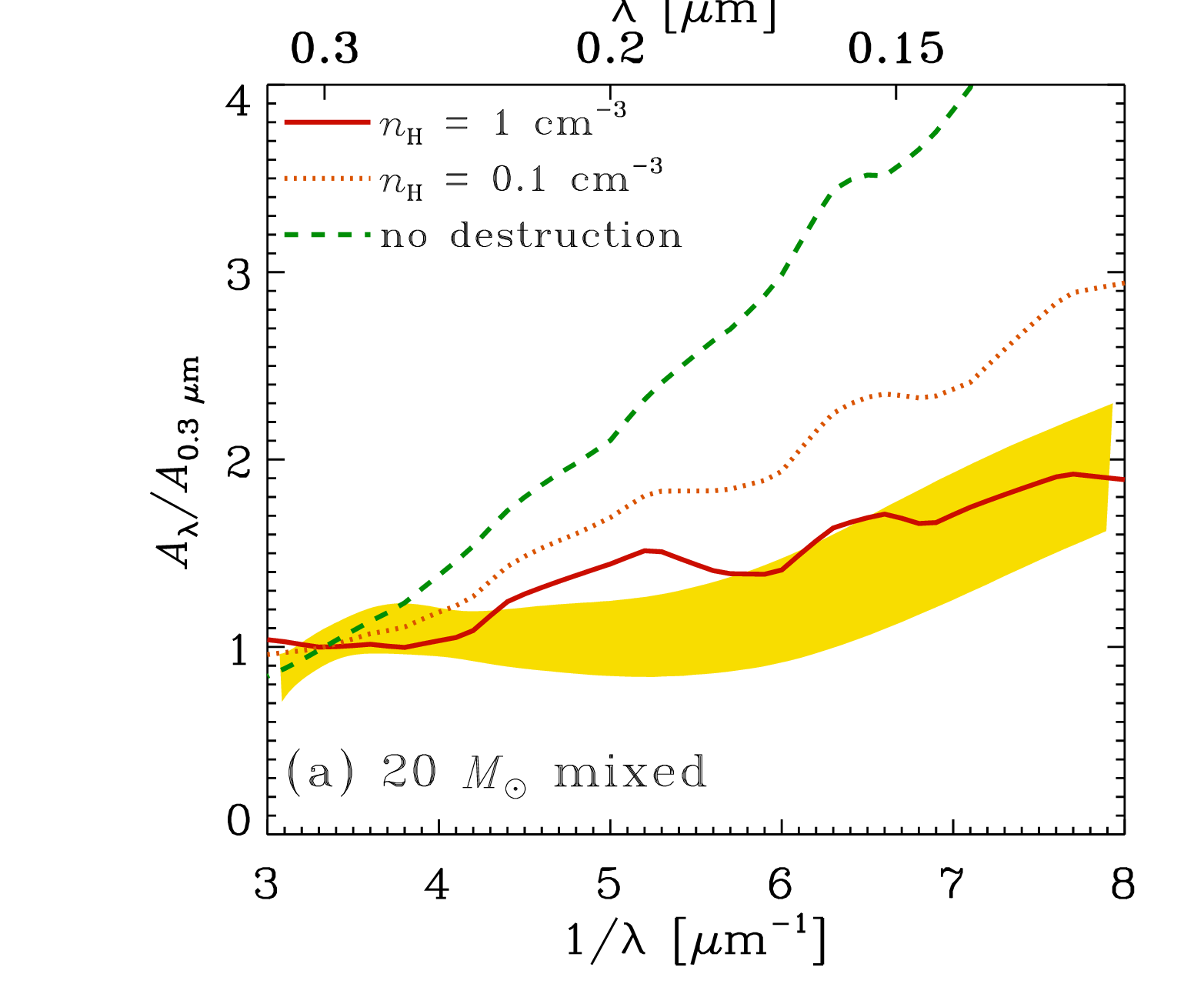}
\includegraphics[width=8cm]{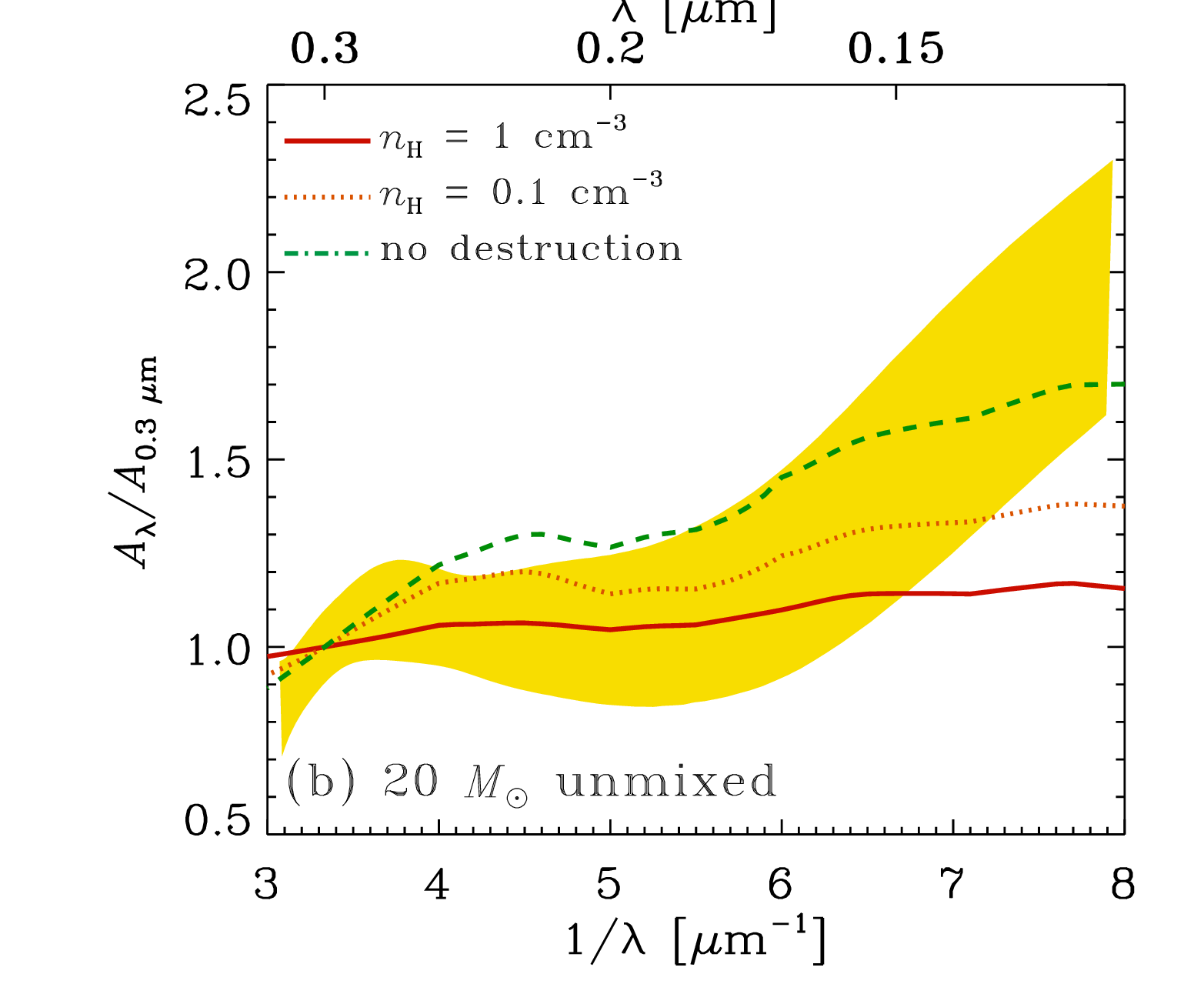}
\includegraphics[width=8cm]{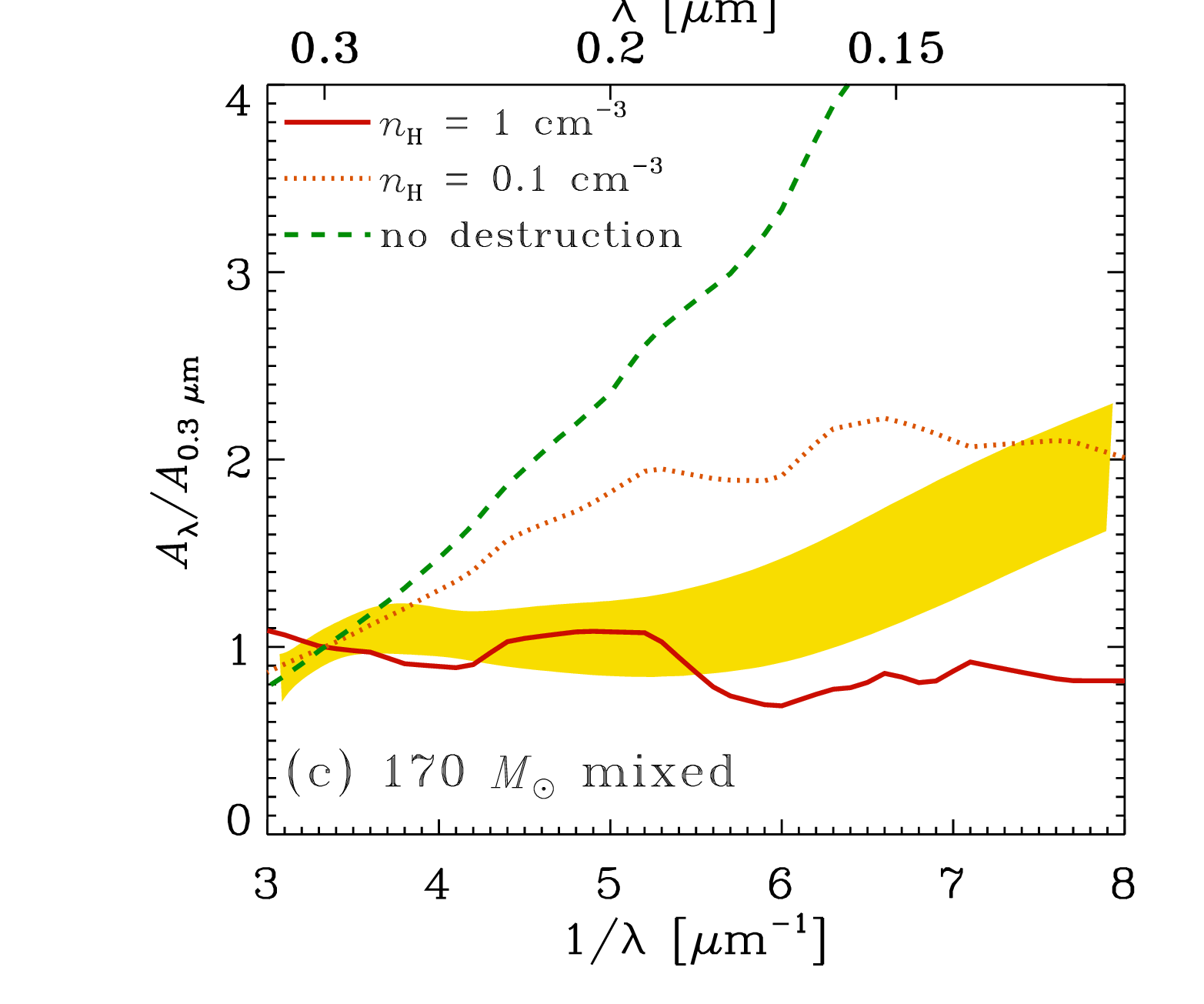}
\includegraphics[width=8cm]{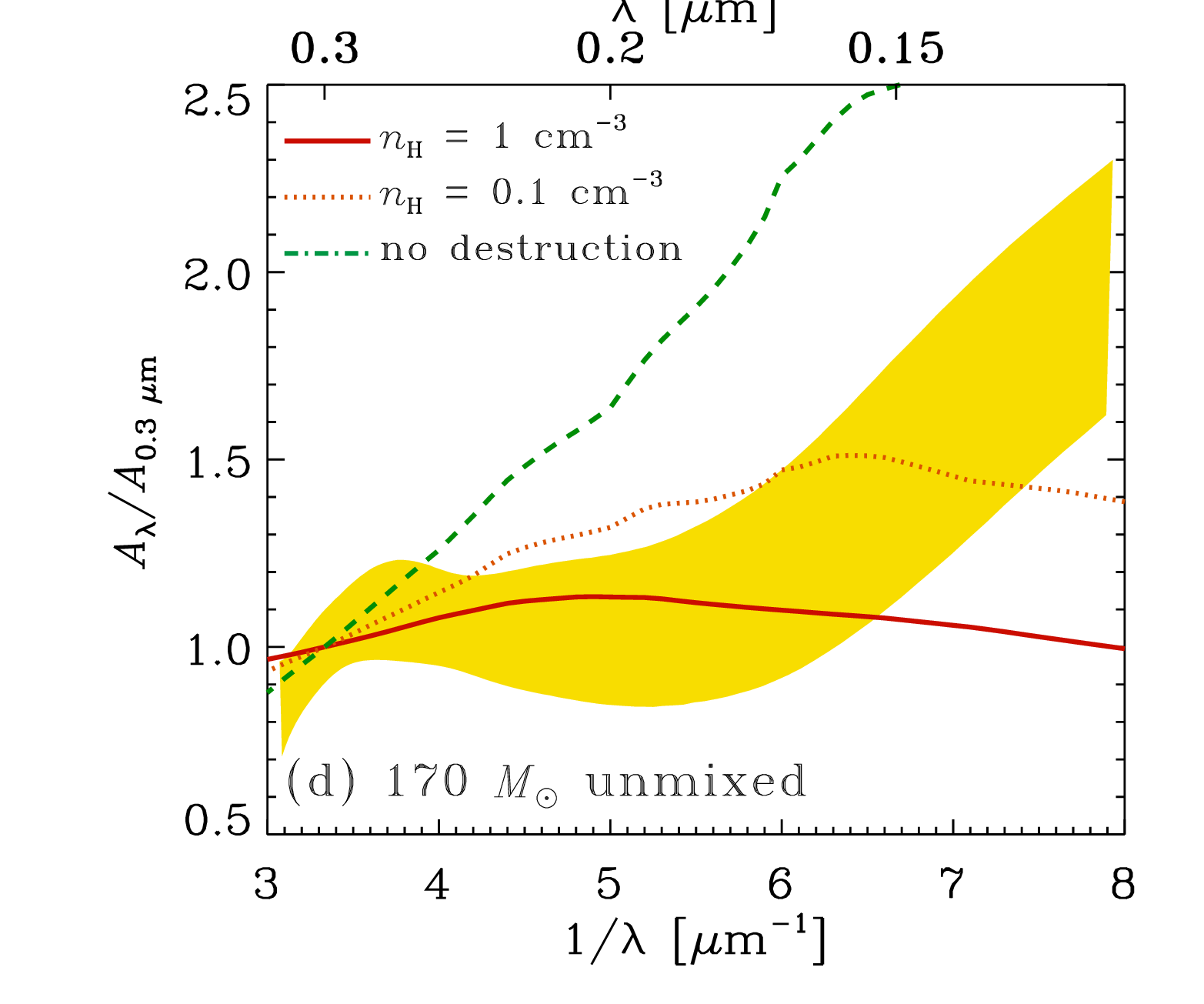}
\caption{Extinction curves normalized at
$\lambda =0.3~\mu$m. The same theoretical
curves as those in Figure \ref{fig:ncomp} are
plotted. The shaded area in each panel show the
range observationally derived by \citet{maiolino04b}
for \sdss\ at $z=6.2$.
\label{fig:maiolino}}
\end{figure*}

H05 show that the dust production model of the unmixed
SNe II in N03 is consistent with the extinction curve
of \sdss. However, it is crucial to compare our new
results including the reverse shock destruction with
\sdss. In particular, we may obtain constraints on the
ambient gas density as well as the progenitors. We
compare our results with the
restframe UV extinction curve of
\sdss. The plausible range derived
by \cite{maiolino04b} is shown by the shaded
areas in Figure \ref{fig:maiolino}.

In each panel of Figure \ref{fig:maiolino}, we show
the theoretical extinction curves calculated for
various ambient hydrogen number densities
(same as Figure \ref{fig:ncomp} but normalized to
the extinction at $\lambda =0.3~\mu$m, i.e.,
$\lambda_0=0.3~\mu$m). We also
show the results of the non-destruction case.
Because of the variation caused by the destruction,
various types of progenitors may be allowed. For
example, the mixed SNe II are consistent with the
observed extinction curve
if $n_{\rm H}\sim 1$ cm$^{-3}$. In all cases,
the ambient number density $n_{\rm H}\ga 1$ cm$^{-3}$
produces too flat extinction curves to be
consistent with the data.

\subsection{Ambient density around SNe}

In the previous subsection, we have shown that the
extinction curves are sensitive to the ambient
density of SNe. Thus, it is important to constrain
the density first. Unfortunately,
there is no direct observational constraint on
the gas density in \sdss.

It is rather interesting to point out that the
extinction curves for $n_{\rm H}\la 0.1$ cm$^{-3}$
of the unmixed SNe II
is quite consistent with the extinction curves of
\sdss. The mixed SNe II also reproduce the
observational extinction curve quite well
if the ambient hydrogen number density is
$\sim 1$ cm$^{-3}$. The extinction curves
of PISNe are too flat for $n_{\rm H}\ga 1$ cm$^{-3}$.
Considering the uncertainty in the observational
data, however, the extinction curves of PISNe are
not rejected for $n_{\rm H}\sim 0.1$--1 cm$^{-3}$.

Theoretically, it is possible to examine the
density evolution around the progenitors of SNe.
\citet{kitayama04} show that in small objects
whose typical halo mass is $\sim 10^6~M_\odot$,
the entire gas cloud is completely swept by the
ionization front, and the dynamical expansion of
the ionized region reduces the ambient density
down to $n_{\rm H}\la 1$ cm$^{-3}$. On the contrary,
in more massive objects, the evacuation of the
gas around a massive star does not efficiently
occur and the final gas density is much higher.
However, if we consider 3-dimensional complex
structure of interstellar medium, the density
around a massive star is hard to predict
theoretically. Thus, at this moment, a direct
constraint on the gas density in
\sdss\ by using for example excitation states
of molecular or atomic lines is desired.

In summary, the current observational extinction
curve at $z\sim 6$ can be explained by the
hypothesis that dust is
produced by SNe II and/or PISNe and destroyed
subsequently by
reverse shocks. In addition, the results in
this paper
newly propose that the ambient density should be
less than $\sim 1$ cm$^{-3}$, since the reverse shock
destruction flattens the extinction curve too much
if $n_{\rm H}\ga 1$ cm$^{-3}$.

\subsection{Flat extinction curves at high $z$?}

As shown
in Section \ref{sec:results}, a flat extinction curve is
naturally produced by reverse shock destruction
in SNe. At $z>5$, where it is probable that SNe II or
PISNe predominantly supply dust grains, it is worth
investigating possibility that
extinction curves are flat.

Indeed, there are some pieces of supporting evidence
for flat extinction curves at $z>5$.
Among the sample in \citet{maiolino04a},
two BAL quasars, SDSS J1044$-$0125 at $z=5.8$
and SDSS J0756+4104 at $z=5.1$, show no reddening but
are detected at
sub-mm \citep{priddey03}. The sub-mm observations
yield dust masses in the range $10^{8-9}~M_\odot$ for
those two quasars. The absence of
reddening and the presence of sub-mm emission
(dust emission) may be
contradictory, but they are consistent if we
assume a flat extinction curve.
An extremely small reddening of a BAL quasar
SDSS 1605$-$0112 at $z=4.9$
\citep{maiolino04a} may also be attributed
to a flat extinction curve.
In the future, extinction curves of high-$z$
sub-mm samples
\citep{bertoldi03,priddey03,robson04,beelen06}
in comparison with theoretical modeling of sub-mm
emission \citep[e.g.][]{takeuchi03,takeuchi05}
enable us to further
investigate the properties and origins of dust
grains in the early Universe.

Also at lower $z$ ($\la 5$), it may be worth to
consider possibility of flat extinction curves.
Direct indications of dust at $z\la 5$ come, for
example, from the
reddening of background quasars \citep{fall89,zuo97}.
The depletion of heavy elements in quasar absorption line
systems, especially damped Ly$\alpha$ clouds (DLAs),
also supports the presence of dust in distant
systems \citep[e.g.][]{pettini94,vladilo02,ledoux03}.
However, there are some observational results that show
no significant reddening of DLAs
\citep{murphy04,ellison05}. The flat extinction
curve proposed above has importance that the lack
of reddening does not necessarily mean the absence of
dust. Other than DLAs, some objects
whose extinction curves are derived by gamma-ray bursts
\citep{chen06} or by gravitational lensing
\citep[e.g.][]{falco99,munoz04} seem to
have a flat extinction curve in UV.

\subsection{Possibility of subsequent steepening}

Dust grains supplied into interstellar spaces suffer
various processes that modify their size distribution.
One of such processes is the destruction by forward shocks
of supernova remnants. The dust destruction by forward
shocks has already been examined by \citet{nozawa06}.
Here we have examined the effect of forward shock
destruction and have confirmed that one passage of
a supernova forward shock has negligible influence on the
extinction curve. This is because after an efficient
sputtering in supernova remnants the opacity is
already dominated by large grains which are hard to destroy.
However, shattering may increase the number of small-size
grains
\citep{borkowski95,jones96}, steepening the extinction curve.
Quantitative study of the effect of shattering on
the extinction curve is left for future work.

\section{CONCLUSION}\label{sec:sum}

We have theoretically investigated the extinction curves
of grains produced in SNe II and PISNe. Since
at high $z(>5)$, low-mass stars cannot be dominant
sources for dust grains, SNe II and PISNe, whose progenitors
are massive stars with short lifetimes, can govern the
dust production. While our previous works (N03 and H05)
did not consider the
reverse shock destruction induced by a collision with
ambient interstellar medium, we adopt the composition
and size distribution of grains of N07, who take
into account the reverse shock destruction.

We have found that the extinction curve is sensitive to
the ambient gas density around SNe, since the efficiency of
reverse shock destruction is largely dependent on it.
The destruction is significant for small-sized grains,
leading to a flat extinction curve in the optical
and UV. Such a large ambient density as
$n_{\rm H}\ga 1$ cm$^{-3}$ produces too flat an extinction
curve to be consistent with the observed extinction curve
for \sdss\ at $z=6.2$.
Although the extinction curve is
highly sensitive to the ambient density, the hypothesis
that the  dust is predominantly formed by SNe at
$z\sim 6$ is still allowed if $n_{\rm H}$ is smaller
than 1 cm$^{-3}$.
For further quantification, the ambient density
should be constrained by some other methods.

It is worth noting that
a flat extinction curve produced by
effective reverse shock destruction may explain the absence
of reddening of systems in which dust is known to be present
by far-infrared/sub-mm emission or by depletion of heavy
elements.

\section*{Acknowledgments}
We thank R. Maiolino and his collaborators for kindly
providing us with their data on the extinction curve
of \sdss. We are also grateful to S. Bianchi for
helpful comments that improved this paper very much.
HH has been supported by Grants-in-Aid for Scientific
Research of the Ministry of Education, Culture, Sports,
Science and Technology (MEXT) of Japan
(Nos.\ 18026002 and 18740097). TN also has been supported
by the same grants (Nos.\ 18104003 and 19740094).
TTT has been supported by the Special Coordination Funds
for Promotion Science and Technology (SCF) commissioned by
the MEXT (MEXT) of Japan.
We fully utilized the
NASA's Astrophysics Data System Abstract Service (ADS).

\appendix

\section{Comparison between Crystal and Amorphous
Grains}
\label{subsec:uncertainty}

As mentioned in Section \ref{subsec:method}, the optical
constants of crystal Si and SiO$_2$ are available,
while we have used those of amorphous Si and SiO$_2$.
Since those species dominate the extinction curves,
it is important to examine the uncertainty caused by
assumed material states.
The optical constants of crystal Si and SiO$_2$ are
taken from \citet{edward85} and \citet{philipp85},
respectively.
As a representative, we examine the case of
$n_{\rm H}=1$ cm$^{-3}$. In the mixed and unmixed cases,
the difference between crystal and amorphous are examined
for the dominant species, SiO$_2$ and Si, respectively.

\begin{figure}
\includegraphics[width=8cm]{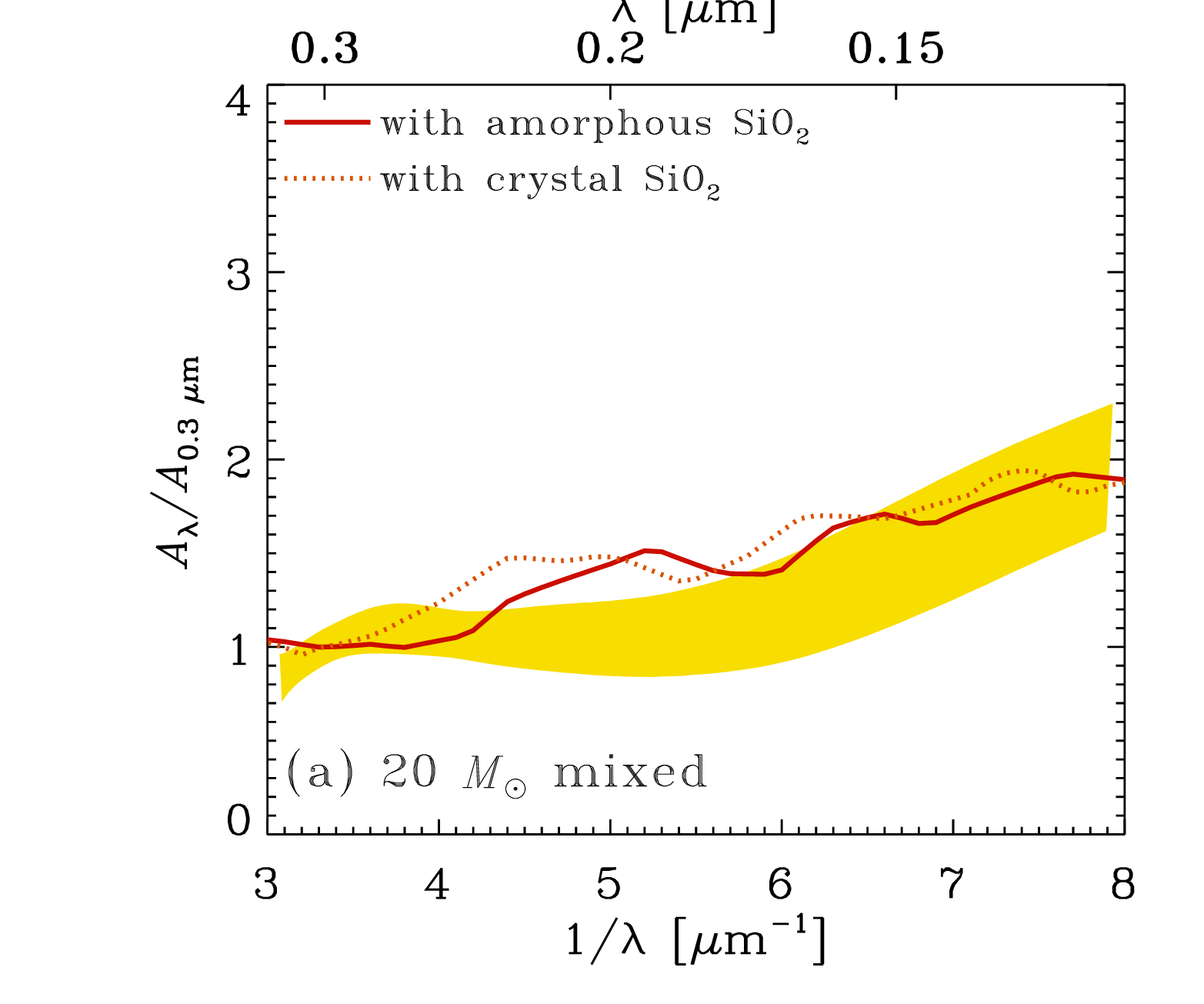}
\includegraphics[width=8cm]{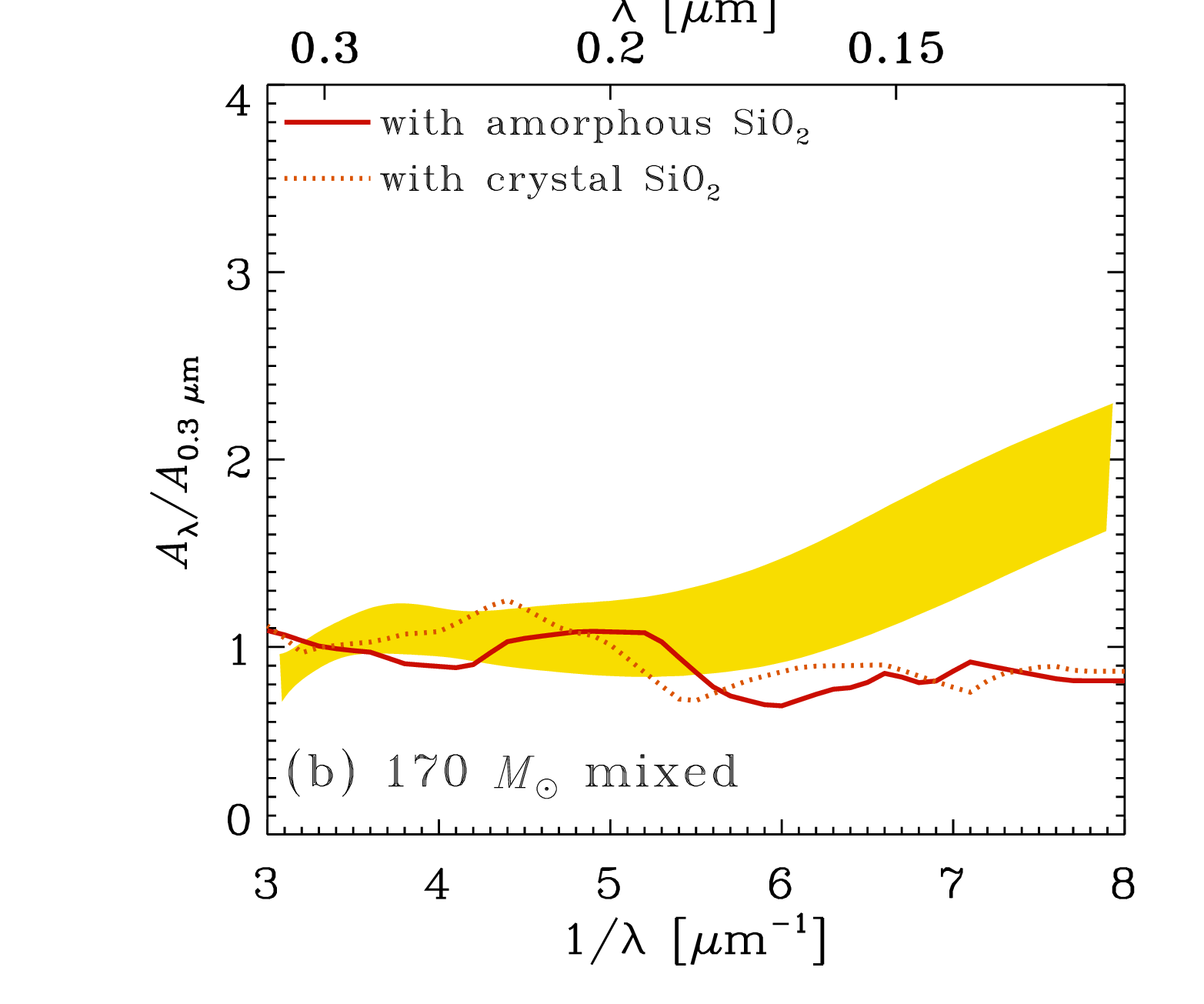}
\caption{The same as Figure \ref{fig:maiolino} for
the mixed SN II and PISN for $n_{\rm H}=1$ cm$^{-3}$
(solid lines in panels a and b, respectively).
 For SiO$_2$, we adopt the optical
constants of amorphous and crystal solids for the
solid and dotted lines. For other species, the same
optical constants as those in Figure \ref{fig:maiolino}
are applied.
\label{fig:sio2}}
\end{figure}

In Figures \ref{fig:sio2}a and b, we show the
extinction curves with the progenitor masses of
20 $M_\odot$ and 170 $M_\odot$, respectively, for the
mixed supernovae. From those figures, we observe that
the overall trend is similar between the two species of
SiO$_2$. Thus we conclude that the two species of
SiO$_2$ are indistinguishable in the extinction curve
within the uncertainty in the current observational data.

\begin{figure}
\includegraphics[width=8cm]{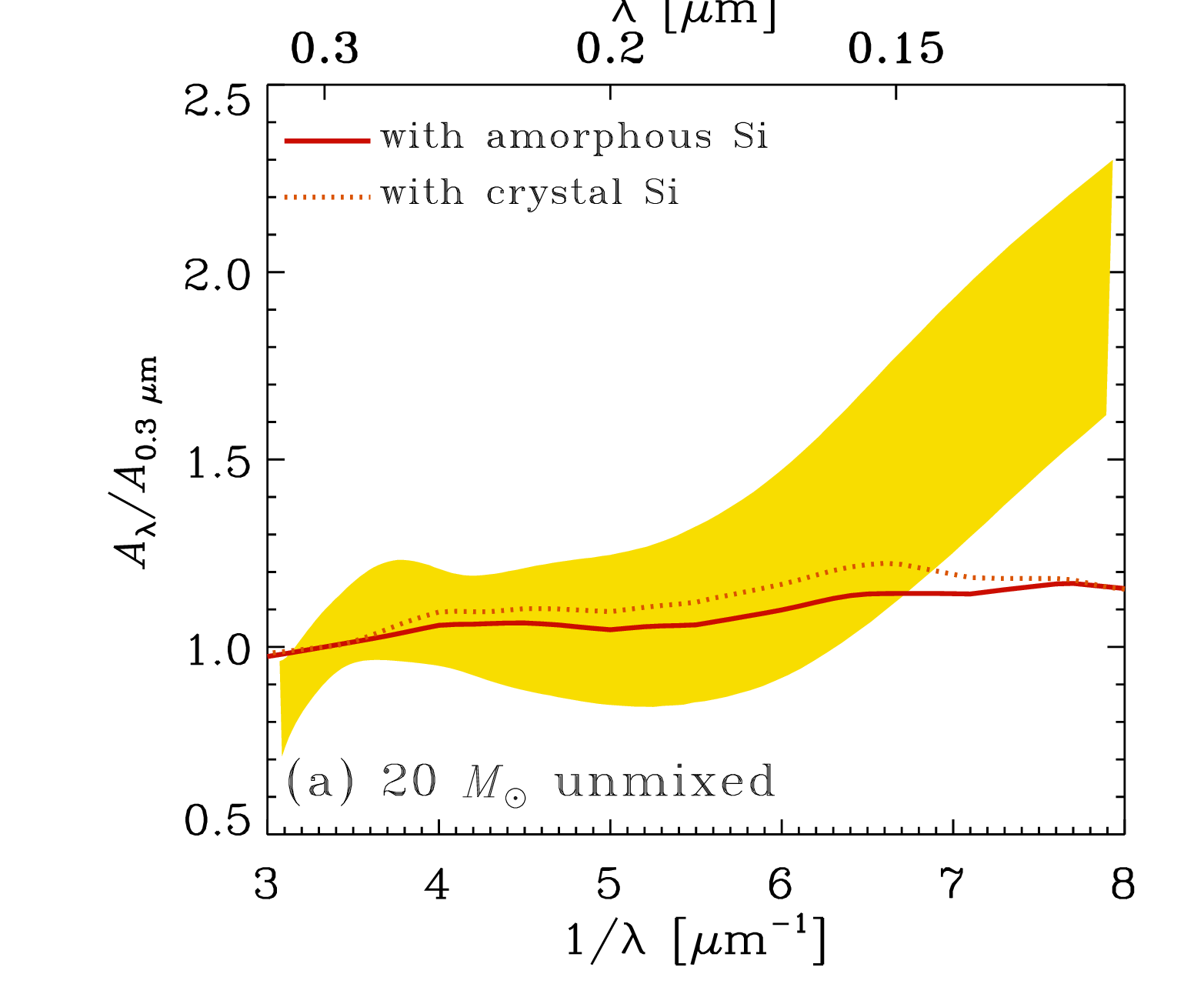}
\includegraphics[width=8cm]{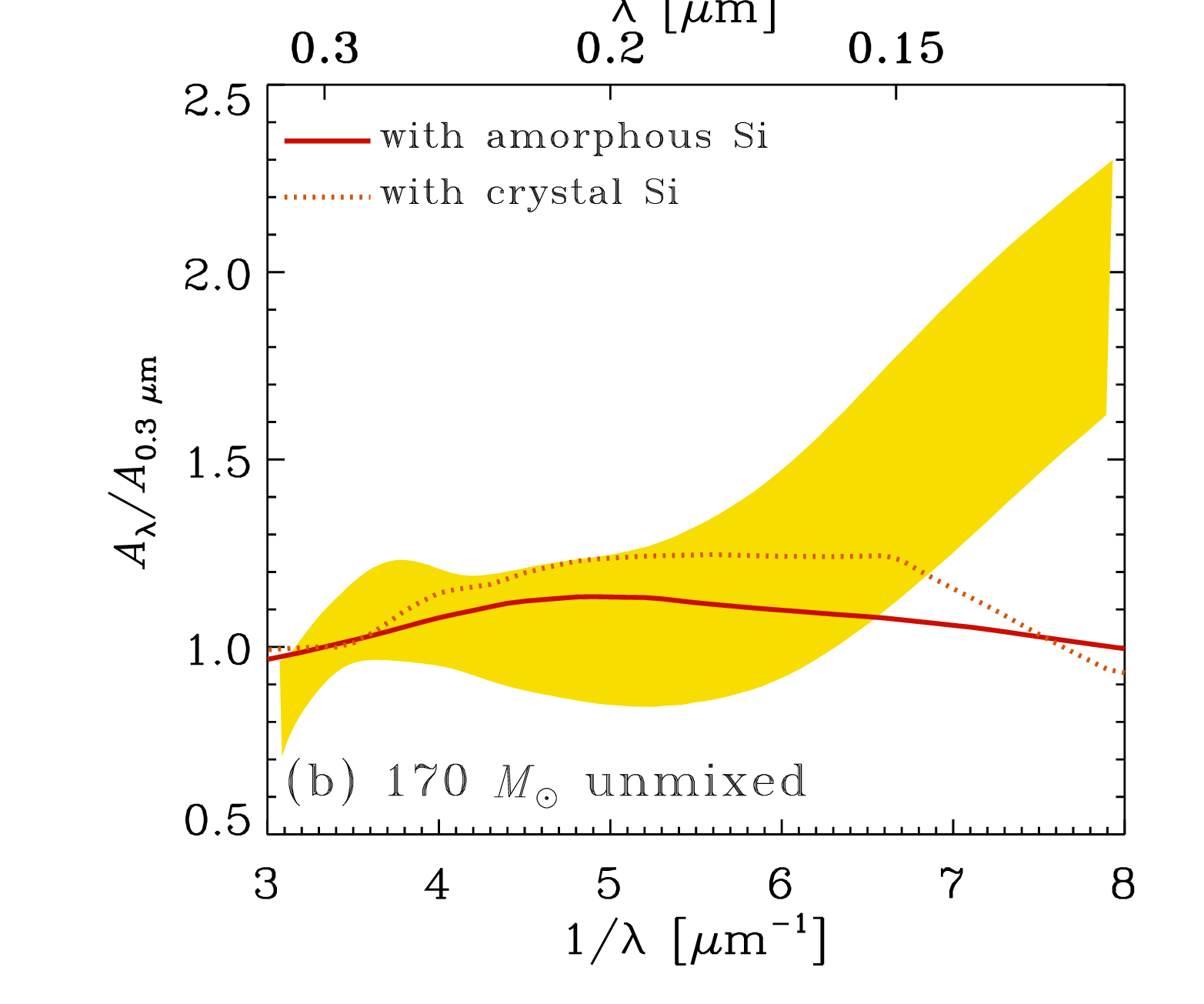}
\caption{The same as Figure \ref{fig:maiolino} for
the unmixed SN II and PISN for $n_{\rm H}=1$ cm$^{-3}$
(solid lines in panels a and b, respectively). For Si,
we adopt the optical
constants of amorphous and crystal solids for the
solid and dotted lines. For other species, the same
optical constants as those in Figure \ref{fig:maiolino}
are applied.
\label{fig:si}}
\end{figure}

In Figure \ref{fig:si}, we make the same kind of
comparison for unmixed SNe, but the difference between
crystal and amorphous species is examined for Si.
As can be seen from the figure, the difference is
smaller than the uncertainty in the observational data.

\end{document}